\documentclass[preprint]{aastex}
\slugcomment{AJ, in press (Sep. 2001)}

\newcommand{\kms}{km~s$^{-1}$}
\newcommand{\subsun}{\mbox{$_{\odot}$}}
\newcommand{\etal}{{\it et al.\/}}
\newcommand{\teff}{$T_{eff}$}
\newcommand{\grav}{log($g$)}

\begin{document}

\title{Abundances in Stars from the Red Giant Branch Tip to the
Near the Main Sequence Turnoff
in M71: I. Sample Selection, Observing Strategy and Stellar Parameters
\altaffilmark{1}}

\author{Judith G. Cohen\altaffilmark{2},  
Bradford B.Behr\altaffilmark{2,3} and Michael M. Briley\altaffilmark{4}}

\altaffiltext{1}{Based on observations obtained at the
W.M. Keck Observatory, which is operated jointly by the California 
Institute of Technology and the University of California}

\altaffiltext{2}{Palomar Observatory, Mail Stop 105-24,
California Institute of Technology}

\altaffiltext{3}{Current address: Department of Astronomy, University of Texas,
Austin, Texas 78712}

\altaffiltext{4}{Department of Physics, University of Wisconsin,
Oshkosh, Wisconsin}

\begin{abstract}

We present the sample for an  abundance analysis of 25 
members of M71 with luminosities ranging from the red giant branch tip to
the upper main sequence.  The spectra are of high dispersion and
of high precision. We describe the observing strategy and
determine the stellar parameters for the sample stars using both
broad band colors and fits of H$\alpha$ profiles. 
The derived stellar parameters agree with those from the
Yale$^2$ stellar evolutionary tracks to within 50 -- 100K for a fixed
\grav, which is within the level of the uncertainties.

\end{abstract}


\section{Introduction}

By virtue of their large populations of coeval stars, the Galactic
globular clusters present us with a unique laboratory for the study of
the evolution of low mass stars.  The combination of their extreme
ages, compositions and dynamics also allows us a glimpse at the early
history of the Milky Way and the processes operating during its
formation. These aspects become even more significant in the context of
the star-to-star light element inhomogeneities found among red giants
in every globular cluster studied to date. The large differences in the surface
abundances of C, N, O, and often Na, Mg, and Al have defied a
comprehensive explanation in the three decades since their discovery.

Proposed origins of the inhomogeneities typically break down into two
scenarios: 1) As C, N, O, Na, Mg, and Al are related to proton capture
processes at CN and CNO-burning temperatures, material cycled through a
region above the H-burning shell in evolving cluster giants may be
brought to the surface with accompanying changes in composition. While
standard models of low mass stars do not predict this ``deep mixing,''
several theoretical mechanisms have been proposed (e.g., the meridional
mixing of Sweigart \& Mengel 1979, and turbulent diffusion, Charbonnel
1994, 1995) with varying degrees of success. Moreover, there is ample
observational evidence that deep mixing does take place during the red
giant branch (RGB) ascent of metal-poor cluster stars 
(see the reviews of Kraft 1994 and
Pinsonneault 1997 and references therein). 2) It has also become
apparent that at least some component of these abundance variations
must be in place before some cluster stars reach the giant branch.
Spectroscopic observations of main sequence turn-off stars in 47 Tuc
(Briley \etal\ 1996, \&
Cannon \etal\ 1998 and references therein), NGC 6752 (Suntzeff \& Smith 1991,
Gratton \etal\ 2001) and most recently
in M71 (Cohen 1999, Briley \& Cohen 2001)
have shown variations in CN
and CH-band and in some cases Na and O line strengths as well 
consistent with patterns found among the evolved giants of these clusters. 

All we know about stellar evolution strongly suggests that
these low mass main sequence
globular cluster stars are incapable of producing significant
amounts of C, N or O while on the main sequence and also incapable
of deep dredge-up.
Both would be required to reproduce the observed abundance variations.
This leads directly to the possibility
that the early cluster material was at least partially inhomogeneous in
these elements or that some form of modification of 
the relative abundances of these elements has
taken place within the cluster since the currently observed
cluster stars were formed. Some suggested culprits include mass-loss
from intermediate mass asymptotic giant branch stars and supernovae
ejecta (see Cannon \etal\ 1998 for an excellent discussion of these
possibilities).

In addition, King \etal\ (1998) have added 
another complication
to the issue of abundance variations within globular clusters.  Their
analysis of a small number of sub-giants in M92 yields
[Fe/H] = $-$2.52 dex
\footnote{The standard nomenclature is adopted; the abundance of
element X is given by [X/H] = log10[N(X)/N(H)] $-$ log10[N(X)/N(H)]\subsun.},
a value smaller by about a factor of two than that derived
from the spectra of giants in M92
by many authors including Cohen (1978) and
Sneden \etal\ (1991).   If this is not due to some error in the
analysis or a variation in non-LTE corrections that has not been
properly included, this result is quite puzzling
since the Fe abundance 
in the photosphere of these stars could not possibly
be altered by mixing.

In an effort to unveil the source of the star-to-star element variations
seen among the light elements within
globular clusters, as well as to determine the constancy, or lack thereof,
of the abundances of the heavy elements such as Fe, where no foreseeable mixing can
be expected to alter its abundance, we have in initiated the present program
to study at high dispersion stars over a large range in luminosity within
the nearer galactic globular clusters.  We begin with the nearest globular cluster
easily reached from a northern hemisphere site, M71.

\section{The Selection of Stars}

Stars were chosen to span the range from the tip of the red giant branch
to the main sequence turnoff of M71.  Membership considerations at this stage
involved location on the existing $B,V$
photometric sequences of Arp \& Hartwick (1971) and, for the more luminous
stars, assignment of a high probability of membership by Cudworth (1985) in
his proper motion survey of this globular cluster.  When  possible, 
stars were chosen which had known
CH and CN band strengths from the survey by Briley, Smith \& Claver (2001) of the
red giant branch or the work of Cohen (1999) for the main sequence
region.  Unpublished spectra of these bands from Cohen were 
available for some of the subgiants as well.  
Within luminosity ranges of 1 mag, an effort
was made to select stars that spanned the full range of observed CH
and CN band strengths, i.e. CN weak and CN strong stars.
Only reasonably isolated stars were selected.

Because this cluster lies at low galactic latitude, field star
contamination is a serious issue.  It was not possible to define
the cluster sequence clearly in the subgiant regime.  There the
evolution is rapid, hence the stellar density along the isochrone is low,
while the number of field stars is rising rapidly towards fainter magnitudes.
Cudworth's (1985) proper motion survey in M71  does not
reach faint enough to include subgiants, and even with the very recent study of 
Geffert \& Maintz (2000), not available at the time our sample
was selected, isolation of a clean sample of subgiants in M71
would be quite difficult.

Throughout this paper, the star names are from Arp \& Hartwick (1971), 
or, when not included in the former, are created from the 
object's J2000 coordinates.

\section{The HIRES Observations}

All spectra were obtained with HIRES (Vogt \etal\ 1994) at 
the Keck Observatory.  A maximum
slit length of 14 arc sec can be used with our instrumental configuration
without orders overlapping.  Since an image rotator for HIRES is available
(built under the leadership of David Tytler), if we can find pairs of program stars
with separations less than 8 arcsec, they can be observed together on
a single exposure.  Ideally pairs consisted of two members 
of the M71 sample, but when that was not feasible, pairs with a random
star of suitable brightness (i.e. brighter than the sample star) were chosen
in the hope that the second star would also turn out to be a member of M71. 

The desired minimum SNR was 75 over a 4 pixel resolution element for
a wavelength near the center of echelle order 56 ($\sim$6400 \AA).
This is calculated strictly from the counts in the object spectrum, and 
excludes noise from cosmic ray hits, sky subtraction, flattening problems, etc. 
Since the nights were dark, sky subtraction is not an issue except at
the specific wavelengths corresponding to strong night sky emission lines, 
such as the Na D doublet.  This SNR goal was
achieved, at considerable cost in observing time, 
for all but the faintest star.  The faintest star
was not a program star, but rather an object that fell within the slit
for a program star setup.  Its SNR is only 50 per resolution element.  
%
%

The  fainter program stars required
integration times of several hours.
If there was more than one potential brighter second star, then both such
could be observed during the course of the exposures for the
fainter star by changing the position angle of the instrument's slit
at some point during the exposure sequence while still keeping the faint M71 star
in the slit. 

Approximate measurements of the radial velocity were made on line,
and if a star was determined to be a non-member, the observations were terminated.
If the probable non-member was the second component in a pair, an attempt
was made to switch to another position angle to pick up a different 
second star, when a possible candidate that was bright enough 
was available within the limits of the 8 arcsec maximum separation.
Through creative use of close pairs, a sample of 29 stars was observed in M71
with HIRES.

To avoid crowding of spectral lines, the observations were centered
at about 6500\AA.  A 1.15 arcsec slit was used, which provides a spectral
resolution of 34,000.  All long integrations were broken 
up into separate exposures, each 1200 sec long, to optimize cosmic
ray removal.

Because the HIRES detector is undersized, our spectra do not cover the 
full length of each echelle order without gaps in the wavelength coverage.
We wanted to include key lines of critical elements,
specifically the 6300, 6363 [OI] lines, the 7770 O triplet, the
Na doublet at 6154, 6160\AA, and the 6696, 6698\AA\ Al I lines.  
Two instrumental configurations were used for the brightest stars,
as it was impossible to create a single one which included all
the desired spectral features in the wavelength range 6000 to 8000 \AA.
In particular, although the 6696, 6698\AA\ Al I doublet is the most useful 
feature of that element in this spectral region, we could not get it
to fit 
into a single instrumental configuration together with the O lines.  
For the faintest stars, only a single configuration was used, which
included the O lines but did not include the Al I doublet.

The spectra were reduced by BBB using Figaro (Shortridge 1993) scripts 
with commands written by McCarthy and Tomaney (McCarthy 1988)
specifically for echelle data reduction. 

Table 1 gives details of the HIRES exposures for each star, with the
total exposure time for the primary and for the Al configuration.
The signal level per pixel in the continuum at 6150 \AA\ is also given, from
which the SNR can be calculated assuming Poisson statistics and ignoring
issues of cosmic ray removal, flattening etc.  The latter become non-negligible
for the very long HIRES integrations necessary as faint as possible
in M71.
Also listed is the radial velocity for each star measured from the HIRES 
spectra and
the probability of membership assigned by Cudworth (1985) based on
his proper motion study, which included only the brighter stars in
the sample.  A montage of spectra of a single echelle
order for selected stars covering the luminosity range of the sample
is shown in Figure~1.

It should be noted that to acquire this set of relatively high precision
and high dispersion spectra took a total of five nights of assigned time
at the Keck Observatory.  One of the assigned nights was used for a
backup program due to poor seeing conditions.

\section{Radial Velocities}

Radial velocities were measured from all the HIRES spectra by
cross correlating orders 56 and 57 against the spectrum
of a bright template star (in practice the brightest observed M71 star) 
from each run.  The radial velocity
of the template star from each run was determined by fitting Gaussians
to 20 strong isolated features in these two orders
using wavelengths from the NIST Atomic Spectra Database Version 2.0 (NIST
Standard Reference Database \#78).
Heliocentric corrections appropriate for each exposure were then applied.

The internal radial velocity errors were calculated following
the precepts of Davis \& Tonry (1979) using the relation 
$\sigma(v_r) = \alpha/[1+R(TD)]$, where the parameter $R(TD)$
is a measure of the ratio of the
height of the peak of the cross correlation to the noise in the
cross correlation function away from the peak.  
The constant $\alpha$ was set at 15 \kms, which represents a value
at the upper end of those found in other recent HIRES programs using similar
instrumental configurations by 
Mateo \etal\ (1998), Cook \etal\ (1999), and C\^ot\'e \etal\ (1999).
Every star except the faintest one observed, 
G53414\_4435,
has an internal error
in $v_r$ under 1 \kms, while this star only has a somewhat larger 
internal error of 1.5 \kms.

Based on their measured radial velocities, four of the 
29 stars observed with HIRES are not members
of M71 (stars Y, G53425\_4608, G53475\_4547 and G53394\_4624).  
All of these four stars were chosen not as members
of the primary M71 sample but as bright(er) stars to complete
a pair, i.e. chosen primarily on the basis of their location 
on the sky without as careful
scrutiny of their colors as was done for the primary sample.

The radial velocities for the 25 members of M71 observed with
HIRES are listed in Table 1.
They have a mean $v_r$ of $-$21.7 \kms.  After removing in quadrature
an internal uncertainty of 1.0 \kms, we find
$\sigma$ = 2.6 \kms.  This is in excellent agreement
with the value of Peterson \& Latham (1986) of
$-$22.5 \kms\ ($\sigma$ = 2.4 \kms) determined from a sample of 17
bright giants in M71. 
 
The only star which shows lines obviously broader than those expected from
the instrumental resolution is M71 I-80, a RHB star.  The profiles
of weak lines of this star suggest that it is rotating at a 
projected velocity of $\sim$8 \kms, an issue which will be 
discussed by B.Behr in a future publication.

\section{The Stellar Parameters}

We seek to determine for the M71 sample of stars the stellar 
atmosphere parameters  necessary
to carry out an abundance analysis from our HIRES data.
We adopt for M71 the distance (3900 pc) and reddening (E(B-V) = 0.25 mag)
from the on-line compilation of Harris (1996).
\footnote{The extinction maps of
Schlegel, Finkbeiner \& Davis (1998)
from their analysis of the COBE/DIRBE database  are not reliable this close to
the galactic plane. For M71, with $b = -4.6^{\circ}$,  
they deduce $E(B-V) = 0.32$ mag.}
The relative extinction in various passbands is taken from
Cohen \etal\ (1981) (see also Schlegel, Finkbeiner \& Davis 1998). 
Based on the high dispersion analysis of Cohen (1983) for
several red giants in M71, since confirmed by 
additional high resolution studies of giants in M71 by
Leep, Oke \& Wallerstein (1987) 
and by Sneden \etal\ (1994),
we adopt as an initial guess a metallicity for the 
cluster of [Fe/H] = $-$0.7 dex.

\subsection{\teff\ From Broad Band Colors - Predictions of the Model Atmospheres}

We utilize here the grid of predicted broad band colors and
bolometric corrections of
Houdashelt, Bell \& Sweigart (2000) based on the
MARCS stellar atmosphere code (Gustafsson \etal\ 1975).  Before proceeding
we demonstrate that the Kurucz and MARCS predicted colors are essentially
identical, at least for the specific colors used here.
 
We compare the colors predicted from the MARCS code from Houdshelt \etal\
with those from
the Kurucz ATLAS code (Kurucz 1992).  We take the predicted $V-K$ color
from each model in the MARCS grid with [Fe/H] = $-0.5$, and interpolate
within the Kurucz color grid at the same abundance and at the \grav\
of the MARCS model
to find  the \teff\ that would be deduced.

A contour plot of the difference $\Delta$\teff(Kurucz - MARCS) 
that
results when the $V-K$ color is used
is shown in Figure 2.  The three contour levels shown
correspond to $\Delta$\teff = 0, 30 and 60 K.  Also shown
in this figure as the thick curve is a 12 Gyr isochrone for M71
from  the very recently completed Yale$^2$ isochrones of
Yi \etal\ (2001).
Along this isochrone, $\Delta$\teff = 0 to 25 K
for the subgiants and main sequence turnoff, and $\Delta$\teff = 
0 to 15 K for the red giant branch.  Throughout the entire displayed
range in \teff,\grav, $\delta$\teff\ ranges from 0 to 50 K.  We thus demonstrate that
to within a tolerance of 25 K, the Kurucz and MARCS temperature scales
from broad band $V-K$ colors are identical.

\subsection{\teff\ From Broad Band Colors}

Broad band $B,V$ colors are available from the work of Arp \& Hartwick (1971).
With such high reddening and metallicity, we chose to ignore the $B$ measurements.
Stetson (2000) provides $V,I$ photometry for about 25\% of the cluster area,
specifically the NE quadrant.  To supplement this, JGC carried out
$V, R, I$ photometry using short exposure images of M71 taken 
with LRIS (Oke \etal\ 1995) at the Keck Observatory
for slitmask alignment purposes.  These frames were
calibrated by observations of standard fields from Landolt (1992).  
The brightest M71 giants were saturated on all the LRIS images.

For a smaller sample of $\sim$235
stars in M71, we have obtained infrared photometry at $K$
using the infrared acquisition and guiding
camera on NIRSPEC (McLean \etal\ 1998, 2000) at the Keck Observatory. 
These were taken during a night dedicated to infrared spectroscopy
in M71, and are basically setup images for the spectroscopy.
The data is calibrated to the standard stars of Persson \etal\ (1998).
A 256 x 256 pixel NICMOS detector is used with a scale of
0.18 arcsec/pixel.  Hence the fields are very small and
main sequence stars dominate the sample.  The
frames were reduced in a standard manner.
This is supplemented by the infrared photometry of Frogel, Persson
\& Cohen (1979) for the upper giant branch of M71. For the
single star in common between the two infrared samples
there is reasonable agreement (see below).  Infrared photometry was obtained
for a few additional stars using the camera of Murphy \etal\ (1995)
at the 1.5m telescope at Palomar Mountain.  With a scale of 0.6
arcsec/pixel and a 256x256 Nicmos array, 
only a small portion of M71 can be covered at once.  Exposures of more than
2000 sec (broken up into many spatial positions and repeats) are
required to reach the fainter stars in our sample, and the crowding
is severe with these relatively large pixels.  However, $K$
magnitudes for five of the stars in the HIRES sample were 
obtained in this way.

The $V,V-K$ color magnitude diagram for M71 for the HIRES sample is shown
in Figure 3.  The different sources of $K$ photometry are indicated by
different symbols.  The stars observed by Frogel, Cohen \& Persson (1979)
are also indicated; three stars from their sample,
whose membership probabilities are below 15\% in the proper motion
study of Cudworth (1985), have been excluded as it is unlikely
that they are members of M71.  There are two stars in the HIRES sample 
with multiple measurements for $K$.  These are indicated by horizontal lines
connecting the relevant points in this figure.  (M71 1-45, the 
brightest star in $V$ in the HIRES sample,
near the tip of the RGB
has two measurements of $K$ differing by only 0.09 mag, difficult
to see on the figure.)

The photometry of Frogel \etal\ (1979) reaches to RHB.  While the number
of stars in common with the previously published photometry is only one,
the consistency of the M71 RGB and RHB delineated by the
published photometry and by our photometry shown in Figure~3 indicates
that our mixing of several different sources for $K$ has been done
in a valid manner.

The observed broad band colors for
each program star ($V-I$ and, when available, $V-K$), corrected for extinction, are used to determine
\teff.  The set of models with metallicity of $-$0.5 dex, nearest to our
initial estimate of [Fe/H], is used.
Table~3 lists the \teff\ deduced from each of $V-I$ and $V-K$, when
available. 

The calibration of our photometric data, as distinct from that of Stetson (2000),
could be better.  We assume an uncertainty of 0.02 mag applies to
$V-I$ from Stetson (2000), an uncertainty of 0.03 mag applies to
colors from the LRIS images, and an uncertainty of 0.05 mag
applies to all $V-K$ colors. Given the relatively high
reddening of M71, there is an additional uncertainty
due to possible spatial variations in reddening across the field of the cluster.
We assume this occurs for $E(B-V$) at a level of 10\%, which is the
fractional variation in $E(B-V)$ detected across much more heavily reddened
globular clusters by Cohen \& Sleeper (1995).
This translates into a total uncertainty in \teff\ of 75 K for giants rising
to 150 K for main sequence stars using $V-I$, divided about equally between
the two contributions (uncertainty in reddening and photometric
colors), and  40 K from $V-K$ for giants rising to 70 K
for dwarfs, with the dominant contribution arising from the
photometric uncertainties.  We adopt the larger of these uncertainties
(those from $V-I$)
as applicable for our \teff\ determinations.

\subsection{Computation of \grav \label{grav}}

Once an initial guess at \teff\ has been established from a broad
band color, it is possible with minimal assumptions
to evaluate \grav\ using  observational data.
The adopted distance modulus, initial
guess at \teff, and an assumed stellar mass (we adopt 0.8 $M$\subsun\
for the upper main sequence stars, and 0.6 $M$\subsun\ for
the RHB stars)  are combined with 
the known interstellar absorption, the predictions of the 
model atmosphere grid
for bolometric corrections as well as a broad band observed $V$ mag to
calculate \grav.

An iterative scheme is used to correct for the small
dependence of the predictions of the model atmosphere grid on
\grav\ itself.  Rapid convergence is achieved.

It is important to note that because of the constraint of
a known distance to M71, the
uncertainty in \grav\ is small, $\le0.1$ dex when comparing
two members of M71.  Propagating an uncertainty of 15\% in the cluster
distance, 5\% in the stellar mass, and 3\% in \teff\ from a
reddening uncertainty of 0.04 mag in $E(B-V)$, and ignoring
any covariance, leads to
a potential systematic error of $\pm$0.2 dex for \grav.

\subsection {\teff\ and \grav\ from H$\alpha$ Profiles}

The profiles of the Balmer lines can also be used in principle to determine 
\teff\ and \grav\ in the temperature range characteristic of the M71 stars.  
The estimates of stellar parameters obtained in this way are
to first order independent of reddening and of any photometric data.
There is little sensitivity to surface gravity or overall abundance.
Given the constraints on \grav\ and known approximate metallicity
imposed by the globular cluster membership of the sample stars, 
the primary dependence of the Balmer line profiles in this regime of
\teff\ and \grav\ for the HIRES sample of members of M71 is on \teff.

We attempt to use the H$\alpha$ profiles for this purpose.
The HIRES spectra themselves are not suitable for this purpose,
as the large scale continuum
determination, particularly in these relatively metal rich cool stars,
is suspect at the level of 1 to 2\% due to imperfect removal of the variation
of the instrumental response across each echelle order.  Instead
the Balmer line profiles were obtained from observations 
with LRIS (Oke \etal\ 1995)
at the Keck Observatory.  A 1200 g/mm grating with 0.7 arcsec 
wide slits was centered at 6500 \AA\ to yield 0.63\AA/pixel or a 
spectral resolution of $\sim$1.7\AA.
The same slitmasks that were designed, fabricated and utilized for the
CH and CN observations of 79 main sequence stars 
in M71 described in Cohen (1999)
were used again in M71 for these observations.  An additional slitmask of
subgiants was also designed and used for this purpose.

The spectra were reduced in the usual fashion using Figaro (Shortridge 1993).
Continuum bandpasses were defined based on examination of the much
higher dispersion HIRES spectra of M71 stars.  The median
value within each of the continuum regions was chosen as a
representative value for the bandpass.  A second order polynomial
fit to the signal for each of the ``line free'' regions was used to 
define the continuum.  To improve the SNR still further 
for the main sequence stars,  the H$\alpha$ spectra of
three to five stars 
of similar luminosity along the
main sequence of M71 were summed, then the resulting profile was normalized.
Figure~4 shows H$\alpha$ profiles for three stars summed 
near the bright end of Cohen's (1999) main sequence sample, which
corresponds in the color-magnitude diagram of M71 to the main
sequence stars in the HIRES sample, as
well as for a subgiant.

These profiles were compared against the spectral flux calculated by 
Hauschildt \etal\ (1999) for [Fe/H] = $-$0.7 dex. 
\footnote {As has already been pointed out by
van't Veer-Menneret \& Megessier (1996), the predicted Balmer line profiles
released with the ATLAS 9 models
of Kurucz (1992) are not valid and fail to reproduce the solar profile.
They found, as do we, that these H$\alpha$ profiles over-estimate
\teff\ by several hundred degrees.}
The grid spacing of the spectral synthesis is 2 \AA.    
We fit the predicted flux of Hauschildt \etal\ in the region of
H$\alpha$ using the same procedure as had been applied to the
stellar spectra to generate a set of model Balmer line profiles
with a normalized continuum level.

Even with the use of
LRIS spectra instead of
echelle spectra, the continuum determination across the
H$\alpha$ profile is still uncertain
by 1\%.  The sensitivity to \teff\ of the wings of the predicted H$\alpha$ profiles 
is not large compared to this potential uncertainty. 
To minimize its effect, we compared the observed
and predicted H$\alpha$ profiles only over
the region  within 3 \AA\ of
the line center.  The resulting values of \teff\ for our M71 stars
are still considerably
higher than those derived from the broad band colors.
Because of the problems mentioned above as well as the potential
impact of continued
small improvements to the broadening theory for Balmer lines
(see, for example, Barklem, Piskunov \& O'Mara 2000), we 
decided to use the H$\alpha$ profiles only to estimate relative
values for \teff\ from star to star within the M71 sample, forcing
agreement with the \teff\ deduced from the colors at the main sequence.

The H$\alpha$ profiles provide measurements of \teff\
which are in agreement with those derived from the stellar colors
to within the uncertainties of each method. 
One might hope to determine the stellar mass at the turnoff
directly from the observations through the gravity dependence of
these Balmer line profiles.  However,
the required precision in the observed Balmer line
profiles of better than 1\% is not easily achieved, nor is it
clear that the theoretical profiles are sufficiently accurate.
Furthermore, the dependence of the Balmer line profiles on
\teff\ is much larger.  Thus 
determining \teff\ itself with sufficient precision to 
then extract a precise value for \grav\
would be extremely difficult.

\section{Comparison of Stellar Parameters with Isochrones}

Table 3 provides a summary of the stellar parameters for the 25 members
of our M71 sample determined both from broad band photometry and from
H$\alpha$ fits.   In addition to the values of \teff\ from 
$V-I$, from $V-K$ and from H$\alpha$ (when appropriate)
\footnote{H$\alpha$ is not included in the determination of \teff\ 
for the three hottest stars
near the M71 main sequence turnoff.},
a mean temperature is listed.  The weight of the 
H$\alpha$ value, when used, is half that
of the values from $V-I$ and from $V-K$.   
With the adopted zero point
for assignment of \teff\ from the H$\alpha$ profiles,  
the good agreement between the three values, consistent with the 
expected observational errors, is gratifying.  

Given that these stars
sample the population of a globular cluster,  \teff\ should
decrease monotonically as the luminosity of the star increases.  
Furthermore stars in the
same region of the cluster isochrone ideally should have very similar stellar
parameters.  The weighted values of \teff\ given in Table~3 do not
quite achieve this.  We therefore slightly adjusted the
weighted \teff\ by not more than 100 K (150 K for star G53392\_4624)
(values typical of our
observational uncertainties) 
while retaining the mean
relationship unaltered to try to achieve this.  
The adopted \teff\ for each star in the M71 sample, listed
in the final column of table~3, is the value
used in the abundance analyses presented in subsequent papers in this series.

For the non-members, 
since their distances are unknown, no value of \grav\ can be obtained
and the derived \teff\
will be incorrect if the reddening is different from the value adopted for M71.

Figure~5 compares the adopted \teff\ and \grav\ for our HIRES sample of
members of M71 with the isochrone predicted for a stellar system
with an age of 12 Gyr with
[Fe/H] = $-0.7$ dex from
the very recently completed Yale$^2$ tracks of Yi \etal\ (2001).
Scaled solar mixture abundances are used in the Yale$^2$ calculations
for all elements heavier than He.

First we note that the set of M71 stars observed with HIRES provides
a reasonable sample of the cluster isochrone from the RGB tip to
the upper main sequence, with the exception of the lack of 
subgiants.

Comparing theory and observation using the set of parameters shown
in Figure~5, quite different from the usual
color-magnitude diagram, is a very stringent test.  
The agreement with 
with the new Yale$^2$ isochrone is quite good.  The \teff\ of the
theoretical giant branch for the metallicity of M71, which is well
known from past work and determined yet again in Paper II,
is only 50 -- 100 K cooler at a fixed \grav\ than are the observed stars.

We already know from comparison with the infrared photometry of 
Frogel \etal\ (1979)
that one cannot ascribe this systematic discrepancy to uncertainties in the 
measurements.  So we now consider the various types of systematic
errors that might have occurred.
There are two known systematic errors in the handling of the observational
data described above.  The first is a systematic underestimate of
\grav\ by 0.04 dex as the theoretically predicted mass 
along the upper RGB is
0.88 M\subsun, while a mass of 0.80 M\subsun was used to calculate the
surface gravities for the cluster stars (excluding the RHB stars)
from the observed magnitudes
and colors.   The second is an underestimate
in \teff\ of less than 20 K because the model grid used to define
the predicted broad band colors had [Fe/H] = $-$0.5 dex, not the
nominal metallicity of M71 of $-$0.7 dex. 

As discussed in \S\ref{grav}, the internal errors from star to star
in \grav\ are small, while the systematic error is dominated
by the uncertainty in the distance, and is indicated in Figure~5
by arrows.  The errors indicated in \teff\ are dominated by
the uncertainty in the reddening.  An overestimate of the
reddening $E(B-V)$ by $\sim$0.04 mag, which seems unlikely, 
could reproduce most of the discrepancy shown in 
Figure~5 through the resulting underestimates of \teff.

Another area of concern is the validity of the
relationships we have adopted between color, \teff, and \grav.
As discussed earlier, we
have carefully checked the consistency of
the predicted colors from Houdashelt \etal\ (2000) with those from 
Kurucz (1992) computed using the ATLAS code, and have also examined
the the comparison with the
empirical color--\teff--[Fe/H] relations for dwarfs and for
giants established by Alonso \etal\ (1996, 1999).  For $V-K$, the 
MARCS and Kurucz predictions are in very close agreement, while the
empirical fits to the angular diameter measurements using the infrared
flux method carried out by Alonso \etal\ yield a \teff\ about 50 K cooler for a fixed
$V-K$ color in the relevant range.

In addition, the theoretical tracks utilized thus far
do not include enhancement of the
$\alpha$-elements, which is common in metal poor globular cluster
giants.  However, the O-enhanced tracks of  
Bergbusch \& VandenBerg (1992) do not fit any better for the 
nominal metallicity of M71.  This is not surprising as 
Bergbusch \& VandenBerg show that to first order the effects of enhancing
O are equivalent to using a model with scaled solar abundances
with an appropriately calculated higher global
metallicity.  This would make
the predicted RGB cooler, making the discrepancy slightly
worse.  Their latest $\alpha$-enhanced models given in
VandenBerg \etal\ (2000) retain this behavior.

We know from Paper II the correct [Fe/H] for M71, and will
shortly know from Paper III the $\alpha$-element enhancements.
With that information plus the stellar parameters of Table~3,
once any small remaining discrepancies between the predicted 
and observed stellar
parameters is understood, one can check for consistency with the new
$\alpha$-enhanced tracks of VandenBerg \etal\ (2000).

The total effect under consideration (i.e. the discrepancy between
the theoretical stellar isochrones and the behavior of the observed
cluster sample in M71 shown in Figure~3)
is only $\sim$50 -- 100 K in \teff.  There are several possible
contributions on the observational side 
which may be large enough to explain it,
including an error in the adopted reddening for the cluster
and uncertainties in the relation utilized between color and stellar
atmospheric parameters.
Hence we have chosen to wait for
more such comparisons to be carried out in the domain of \teff, \grav\  
for
additional clusters in future papers before speculating further on this issue.

\section{Looking Forward}

With this information in hand, we are ready
to carry out an abundance analysis based on measurements
of equivalent widths from the HIRES spectra of the M71 sample.
An analysis of the Fe abundances for this
sample of M71 stars is presented in the next paper 
in this series (Ram\'{\i}rez \etal\ 2001).

\acknowledgements

The entire Keck/HIRES and LRIS user communities owes a huge debt to 
Jerry Nelson, Gerry Smith, Steve Vogt, Bev Oke, and many other 
people who have worked to make the
Keck Telescope and HIRES and LRIS a reality and to operate and 
maintain the Keck Observatory. We are grateful to the 
W. M.  Keck Foundation for the vision to fund
the construction of the W. M. Keck Observatory.
We thank Peter Stetson for supplying his M71 photometry in easily
accessible form and Peter Hauschildt for calculating a grid
of H$\alpha$ profiles for us. Partial support  
was provided to MMB by a Theodore Dunham, Jr. grant
for Research in Astronomy and by the National Science Foundation under
grants AST-9819614 to JGC and AST-9624680 to MMB.

\clearpage

%
%
\begin{deluxetable}{lrrrrrrrr}
\tablenum{1}
\tablewidth{0pt}
\tablecaption{The Sample of Stars in M71}
\label{tab1}
\tablehead{\colhead{ID\tablenotemark{a}} & \colhead{$V$} & 
\colhead{Date Obs.} & \colhead{Primary} & \colhead{Al.}
& \colhead{Signal/pixel}  
& \colhead{$v_r$} & \colhead{$\mu$ Prob.\tablenotemark{b}} 
& \colhead{Notes} \nl 
\colhead{} & \colhead{(mag)} & \colhead{} & \colhead{(sec)} & \colhead{(sec)} 
& \colhead{(DN)\tablenotemark{c}} &
\colhead{(km s$^{-1}$)} & \colhead{(\%)} \nl
}
\startdata
1-45 & 12.36 &   Aug.1999 &  400 & 300   & 2425 & $-$19.0 & 99 \nl
I    & 12.42 &   Aug.1999 &  600 & 400   & 2495 & $-$13.9 & 99 \nl
1-66 & 13.07 &   Aug.1999 & 1000 & 400   & 3325 & $-$23.6 & 96 \nl
1-64 & 13.12 &   Aug.1999 &  700 & 1200  & 1535 & $-$17.1 & 99 \nl
1-56 & 13.21 &   Aug.1999 &  700 & ...   &  770 & $-$21.2 & 99 \nl
1-95 & 13.35 &   Aug.1999 & 1800 & 900   & 3565 & $-$20.3 & 99 \nl
1-81 & 13.68 &   Aug.1999 & 1000 & 400   & 1750 & $-$24.3 & 99 \nl
Y    & 13.95 &   Aug.1999 & 3600 & ...   & 8800 & $-$2.1~ & 0 &  d \nl
1-1  & 14.14 &   Aug.1999 & 600 & 300    &  675 & $-$23.5 & 99 \nl
1-80 & 14.45 &  June 2000    & 900 & ... & 4690 & $-$20.7 & 99 & HB\nl
1-87 & 14.47 &  June 2000   & 2100 & 900 & 2940 & $-$22.8 & ...&  HB \nl
1-94 & 14.58 &  Aug.1999 & 1800 & 900    &  975 & $-$24.6 & 98 & HB \nl
1-60 & 14.55 &  Aug.1999 & 1800 & ...    & 1015 & $-$20.8 & 99 \nl
1-59 & 14.71 &  Aug.1999 & 1800 & ...   & 875 & $-$24.1 & ... \nl
G53476\_4543 & 15.07 & Aug.1999 & 7200 & ...  & 4300 & $-$22.5 & ... \nl
2-160 & 15.14 &  June 2000  & 2100 & 900 & 1600 & $-$25.2 & 5 \nl
G53447\_4707 & 15.16 &  Aug.1999 & 7200 & 1200 & 3575  & $-$19.8 & 28 \nl
G53425\_4608 & 15.47 &  Aug.1999 & 1200 & 1200 &  780 & +16.5 & 0 & d \nl
G53445\_4647 & 15.59 &  June 2000  & 3600 & ... & 865 & $-$19.2  &  88 \nl
G53447\_4703 & 16.03 &  Aug.1999 & 7200 & 1200  & 1515 & $-$27.2 \nl
G53425\_4612 & 16.32 &  Aug.1999 & 1200 & 1200  & 335 & $-$21.0 \nl
G53477\_4539 & 16.33 &  Aug.1999 & 7200 & ...  & 1255 & $-$11.4  \nl
G53475\_4547 & 16.63 &  Aug.1999 & 3600 & ...  & 600 & +36.3 &  & d \nl
G53457\_4709 & 16.75 &  June 2000   & 4500 & ...   & 2960 & $-$21.1 \nl
G53391\_4628 & 16.86 &  Aug.1999 & 7200 & ...   & 1210 & $-$21.3 \nl
G53394\_4624 & 16.95 &  Aug.1999 & 4800 & ...   & 700 & +3.4 &  & d \nl
G53417\_4431 & 17.60 &  Aug.1999 & 12000 & ...  & 875 & $-$19.8 \nl
G53392\_4624 & 17.72 &  Aug.1999 & 12000 & ...  & 820 & $-$22.1 \nl
G53414\_4435 & 17.97 &  Aug.1999 & 12000 & ...  & 425 & $-$21.4 \nl
%
%
\enddata
\tablenotetext{a}{Identifications are from Arp \& Hartwick (1971)
or are assigned based on the J2000 coordinates, rh rm rs.s dd dm dd becoming
Grmrss\_dmdd.}
\tablenotetext{b}{This is the probability of membership assigned
by Cudworth (1985) on the basis of his proper motion survey.}
\tablenotetext{c}{The CCD gain is 2.4 e/DN.  The signal is measured in
the continuum of the spectra taken with the primary HIRES configuration
near 6150\AA.}
\tablenotetext{d}{This star is presumed to not be a member of M71.}
\end{deluxetable}

%
%
\begin{deluxetable}{llllrrr}
\tablenum{2}
\tablewidth{0pt}
\tablecaption{Photometry for the M71 Sample}
\label{tab2}
\tablehead{\colhead{ID\tablenotemark{a}} & \colhead{$V$} & 
\colhead{$R$} & \colhead{$I$} & \colhead{$K$} 
& \colhead{RA} & \colhead{Dec}  \nl
\colhead{} & \colhead{(mag)} & \colhead{(mag)} & \colhead{(mag)} & \colhead{(mag)} 
& \colhead{(J2000)}  \nl
}
\startdata
\multispan{3}{Probable Members:} \nl
1--45 & 12.36 &  ...   &  ...  &    8.12  &   19 53 48.37 &   +18 48 00.3  \nl
I    & 12.423 &  ...   & 10.76 &   8.56 &   19 53 44.74 &  +18 46 35.1   \nl
1--66 & 13.071 & ...  &  11.49 &   ...  &    19 53 45.22 &  +18 46 55.5   \nl
1--64 & 13.122  & ...  &  11.49  &  9.32  &  19 53 46.12  & +18 47 26.2   \nl
1--56 & 13.21 &  ...  &  11.77 &   9.79  &   19 53 48.40  & +18 48 23.5 \nl
1-95 & 13.35  & ...  &  11.94 &    ...  &    19 53 41.01  &  +18 46 04.8   \nl
1--81 & 13.68 & ...  &   12.184
 &  10.22  &   19 53 45.48 &  +18 46 49.7   \nl
1--1  & 14.14  & ... &   12.76 &    10.92  &   19 53 52.32 &  +18 44 52.9   \nl
1--80 & 14.45  & 13.85 & 13.29 &    ...  &    19 53 44.20 &  +18 46 48.0  \nl 
1--87 & 14.47 &  13.83 &  13.28 &    ...  &   19 53 45.58 &  +18 45 48.9 \nl
1--94 & 14.58 &  13.94 & 13.42 &   12.01 &     19 53 40.85  &  +18 46 00.9  \nl
1--60 & 14.55 & 13.83  & 13.23  &  11.69  &   19 53 41.78  &  +18 48 41.5   \nl
1--59 & 14.71 & 13.92 & 13.22 &    11.39  &  19 53 42.06 &   +18 48 37.3   \nl
G53476\_4543 & 15.07  & 14.38  & 13.77  &  ... & 19 53 47.62 &  +18 45 43.2 \nl
2--160 & 15.14  & 14.46 & 13.89 &   ...  &    19 53 45.19 &   +18 48 33.2  \nl
G53447\_4707 & 15.16 & 14.48 & 13.94 &  12.50  &  19 53 44.65  & +18 47 07.4 \nl
G53445\_4647 & 15.59 &  14.88 & 14.31\tablenotemark{b} &   ... &  19 53 44.50  & +18 46 47.0   \nl
G53447\_4703 & 16.03 &  15.27 & 14.74 & 13.20  &  19 53 44.65 &  +18 47 03.3 \nl
G53425\_4612 & 16.32  & 15.66 & 15.13  &  ... &  19 53 42.45 &  +18 46 11.7 \nl
G53477\_4539 & 16.33  & 15.66 & 15.10 &  ...  &   19 53 47.72  & +18 45 39.2  \nl
G53457\_4709 & 16.75 &  16.02 & 15.57 &  ...  &   19 53 45.69 &  +18 47 08.8 \nl 
G53391\_4628 & 16.86 &  16.19 & 15.60  &  ... &   19 53 39.05 &  +18 46 28.2 \nl
G53417\_4431 & 17.60 &  17.05 & 16.52 &   15.49 &  19 53 41.72 &  +18 44 31.2 \nl
G53392\_4624 & 17.72 &  17.13 & 16.66  &  ... & 19 53 39.18 &  +18 46 23.9 \nl
G53414\_4435 & 17.97 &  17.44 &  ...  &    15.78 &  19 53 41.37 &  +18 44 34.8  \nl
~ \nl
\multispan{3}{Probable Non-Members:} \nl
Y    & 13.95 &  ...  &  12.62 &   11.13 &    19 53 41.31 &  +18 44 27.4   \nl 
G53425\_4608 & 15.47 &  14.68 & 14.08 &   ...  &    19 53 42.48 &  +18 46 07.7 \nl
G53475\_4547 & 16.63 &  16.13 & 15.80 &   ...  &  19 53 47.53  & +18 45 47.3 \nl
G53394\_4624 & 16.95 &  16.35 & 15.91 &   ... &   19 53 39.41 &  +18 46 23.8 \nl
\enddata
\tablenotetext{a}{Identifications are from Arp \& Hartwick (1971)
or are assigned based on the J2000 coordinates, rh rm rs.s dd dm dd becoming
Grmrss\_dmdd.}
\tablenotetext{b}{This star has $V, I$ from Stetson as well as from the
short LRIS images, but Stetson's 
$I$ is very discrepant from that from the LRIS short images.  This is
the only such case found so far.}
\end{deluxetable}

%
%
\begin{deluxetable}{lllllcl}
\tablenum{3}
\tablewidth{0pt}
\tablecaption{Stellar Parameters for the M71 Sample}
\label{tab3}
\tablehead{\colhead{ID\tablenotemark{a}} & 
\colhead{\teff\ (K)} & \colhead{\teff\ (K)} & \colhead{\grav} & 
\colhead{\teff\ (K)} &
\colhead{\teff (K)} & \colhead{\teff(K)} \nl 
\colhead{} & \colhead{($V-K$)} & \colhead{($V-I$)} & \colhead{} & 
\colhead{(H$\alpha$)} & 
\colhead{(Weighted)\tablenotemark{b}} & \colhead{(Adopt)}   \nl
}
\startdata
1--45      &  3950   &  ...  &  0.9 &    ... &      3950    &  3950 \nl
I         & 4120  &  4175  &  1.0  &   ...  &     4150  &    4150 \nl
1--66      &   ...  &    4305 &   1.35 &   ...   &    4310   &   4250 \nl
1--64   &      4170  &  4225  &  1.35  &  ...  &    4200 &     4200 \nl
1--56    &     4475  &  4560  &  1.6  &   ...   &    4525   &   4525 \nl
1--95    &    ...   &  4630  &  1.65 &   ...  &     4630  &    4550 \nl
1--81    &     4450  &  4460  &  1.75 &   ...   &    4455  &    4550 \nl
1--1     &   4625  &  4710  &  2.05 &   ...  &     4670 &     4700 \nl
1--80\tablenotemark{c}    &     ...  &  5290  &  2.45 &   ...  &     
5290  &   5300  \nl
1--87\tablenotemark{c}    &  ...  &  5205 &  2.45  & ...  &  5205 &  5300 \nl
1--94\tablenotemark{c} &    5320  &   5290  &  2.45  &  ...  &  5315 &  5300 \nl
1--60   &  4980  &  4845  &  2.3  &   ...  &     4910 &  4900 \nl
1--59    &  4560 &   4480 &   2.3   &  ...  &      4520  &    4600 \nl
G53476\_4543 & ...  &  4890  &  2.65  &  ...  &   4890  &    4900  \nl
2--160  &    ...  &   5175 &   2.7 &    ... &    5175 &   5100  \nl
G53447\_4707 &  5225  &  5130 &   2.75  &  5300   &   5200 &   5175 \nl
G53445\_4647 & ...  &   4960 &   2.85 &   ...  &     4960  &    5050 \nl
G53447\_4703 & 5035 & 4920 & 3.0 & ... & 4985 & 5000 \nl
G53425\_4612 & 5090  &  5205  &  3.15  &  5100  &    5140   &   5150 \nl
G53477\_4539 &  ...  &   5090 &   3.15 &  5300   &   5160   &   5150 \nl
G53457\_4709  & ...  &   5240 &   3.35  &  5400  &    5290  &    5200 \nl
G53391\_4628 &  ...  &  5010 &   3.35 &   ...   &    5010  &    5100 \nl
G53417\_4431 &  6010 &  5580  &  4.05 &   5800\tablenotemark{d}  &  5800 &     5800 \nl
G53392\_4624 &  ...  &  5650  &  4.05  &  5800\tablenotemark{d}  &  5650   &   5800 \nl
G53414\_4435 &  5895 &  ...   &   4.15 &   5800\tablenotemark{d}  &  5895   &   5900 \nl
\enddata
\tablenotetext{a}{Identifications are from Arp \& Hartwick (1971)
or are assigned based on the J2000 coordinates, rh rm rs.s dd dm dd becoming
Grmrss\_dmdd.}
\tablenotetext{b}{\teff\ from the H$\alpha$ profile has half weight.}
\tablenotetext{c}{These are RHB stars in M71.}
\tablenotetext{d}{The H$\alpha$ profiles are not used to derive \teff\
for these stars.}
\end{deluxetable}


\clearpage

\begin{figure}
\epsscale{0.7}
\plotone{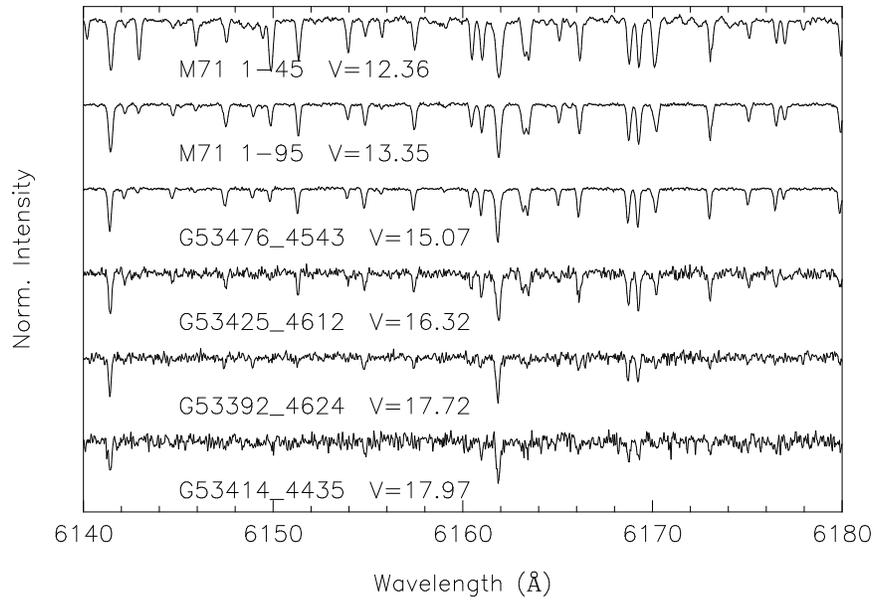}
\caption[figure1.ps]{A section of order 58 is shown for the brightest (at $V$)
M71 star in our sample at the top of the figure and the faintest at the
bottom.  Starting with the brightest star, we display stars in
the sample in increments of five in order of decreasing luminosity,
omitting the RHB stars.
\label{fig1}}
\end{figure}

\begin{figure}
\epsscale{0.7}
\plotone{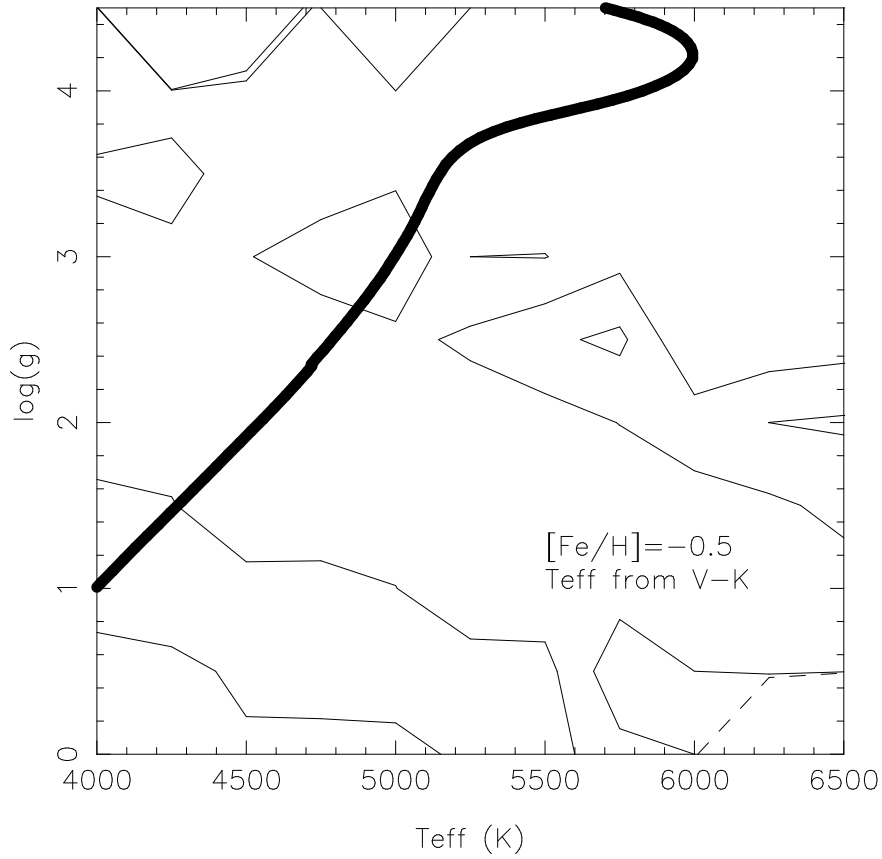}
\caption[figure2.ps]{To demonstrate that the Kurucz and MARCS
grids of predicted colors are essentially identical, 
contours of $\Delta$\teff\ computed for $V-K$ are
displayed. $\Delta$\teff\ is the difference between the \teff\ predicted
from the ATLAS models of Kurucz (1992) and the MARCS models of Houdshelt, 
Bell \& Sweigart (2000) for a fixed abundance ([Fe/H] = $-$0.5 dex) and
a $V-K$ color taken from the MARCS grid.  The contour levels are
0, 30, and 60 K.  The thick curve is a 12 Gyr isochrone for M71 taken
from the Yale$^2$ tracks of Yi \etal\ (2001).  See the text for details.
\label{fig2}}
\end{figure}

\begin{figure}
\epsscale{0.7}
\plotone{figure3.ps}
\caption[figure3.ps]{The $V,~V-K$ color-magnitude diagram for M71 for the
HIRES sample.  Filled circles denote measurements
from SCAM/NIRSPEC, open circles denote measurements from the P60 IR camera,
and crosses denote the set of RGB and RHB stars from Frogel, Persson \& 
Cohen (1979).  For each of the two stars with more than one independent
observation at $K$, horizontal lines connect the pair of points.
\label{fig3}}
\end{figure}

\begin{figure}
\epsscale{1.0}
\plotone{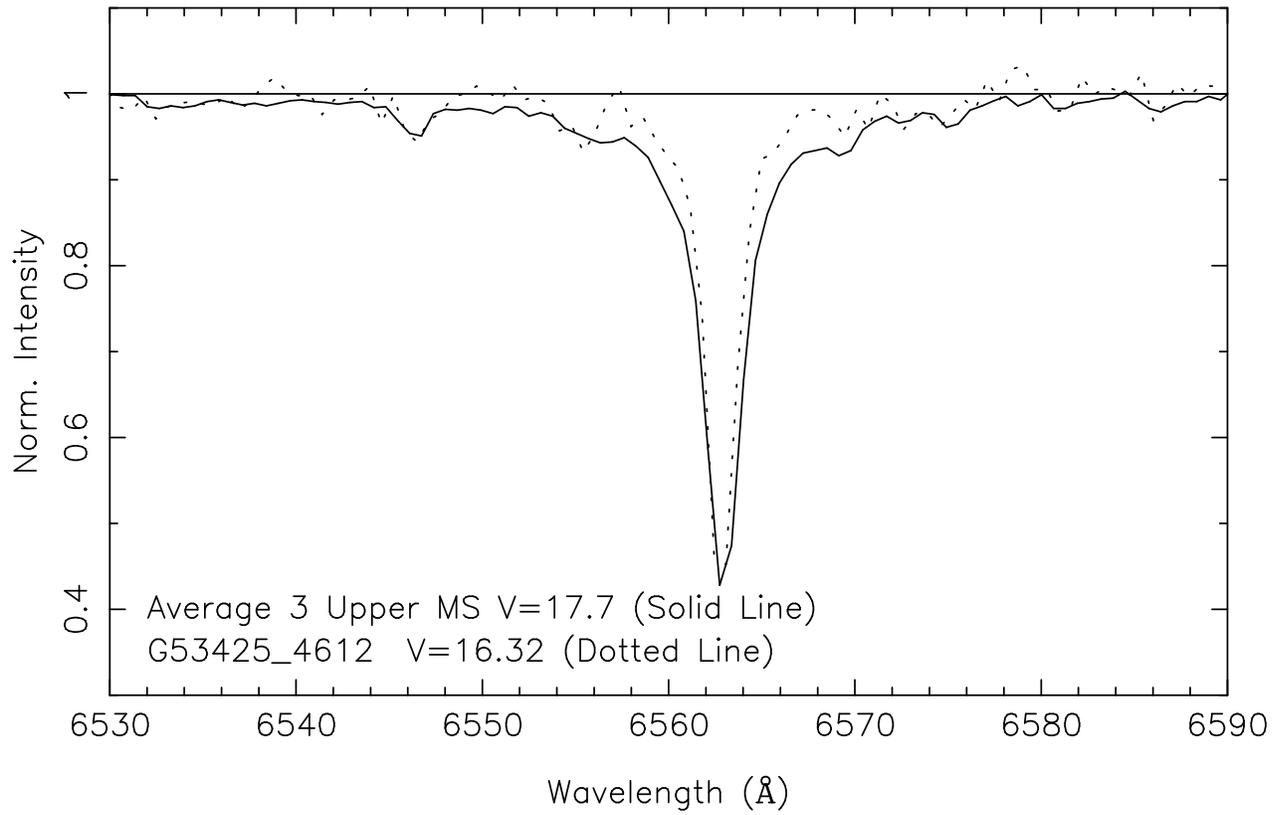}
\caption[figure4.ps]{Profiles of H$\alpha$ from LRIS spectra
are shown for a subgiant
and for the average of three main sequence stars near the bright end of
the sample of Cohen (1999).
\label{fig4}}
\end{figure}

\begin{figure}
\epsscale{0.8}
\plotone{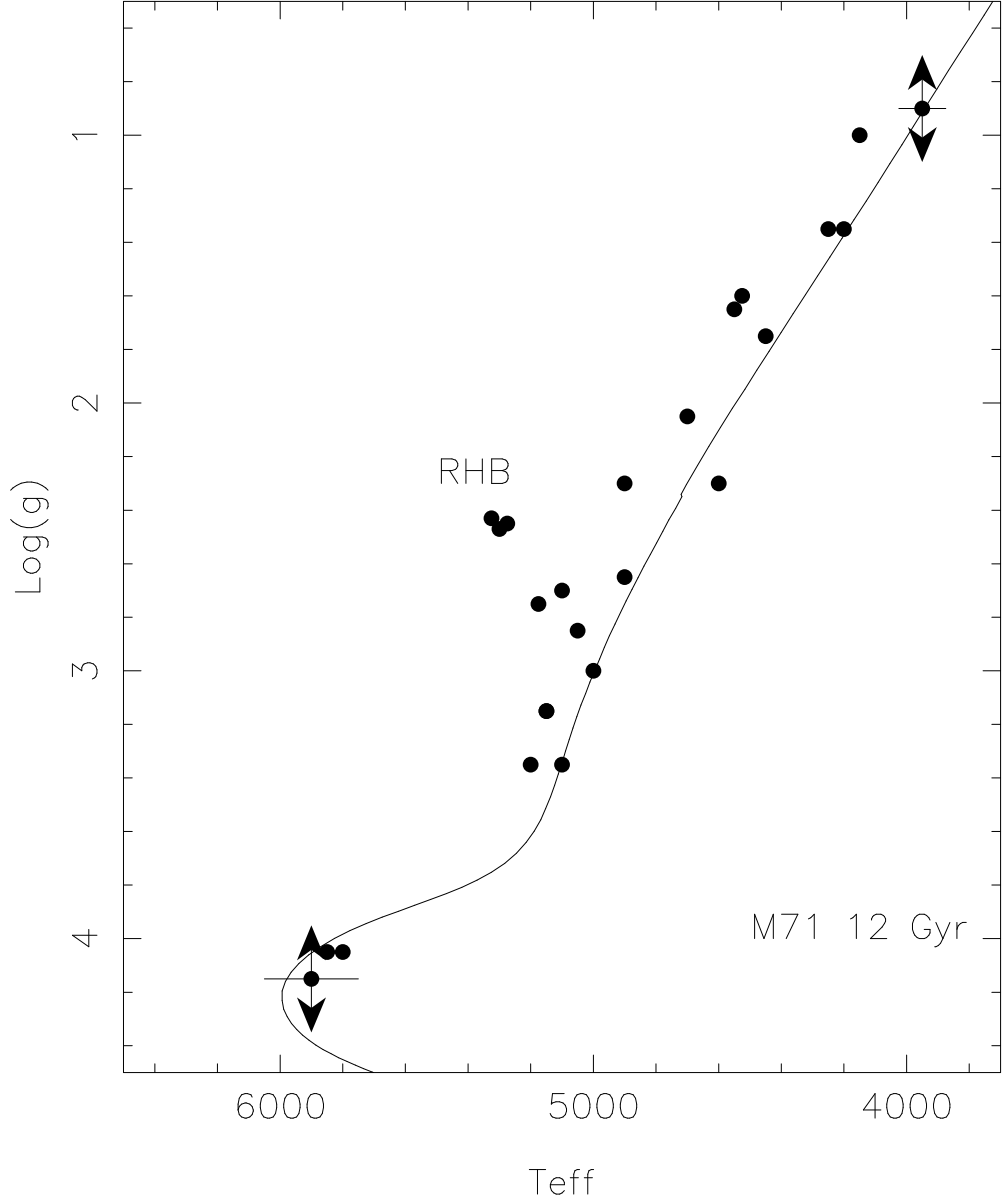}
\caption[figure5.ps]{The \teff\ and \grav\ deduced here for the sample
of M71 members with HIRES spectra is shown as is the 
12 Gyr Yale$^2$ isochrone of Yi \etal\ (2001) for
[Fe/H = $-0.7$ dex (solid curve).  A distance of 3900 pc with a reddening
$E(B-V) = 0.25$ mag has been adopted.   The arrows in \grav\
indicate the systematic error which is dominated by the contribution
from the distance uncertainty; the internal error from star to
star within M71 is considerably smaller. The error bars 
in \teff\ shown for
the most and least luminous M71 stars in the HIRES sample are dominated by uncertainties in the reddening and are typical of the sample.

\label{fig5}}
\end{figure}

\end{document}